\documentclass{moriond}

\def\XibzstarstarA    {$\Xi_b{(6087)}^{0}$}

\def\XibzstarstarB    {$\Xi_b{(6095)}^{0}$}
\def\XibmstarstarB    {$\Xi_b{(6100)}^{-}$}
\def\Xibsigned        {$\Xi^{(-,0)}_b$}

\def\Xibzprime        {${\Xi'}^{0}_b$}
\def\Xibmprime        {$\Xi'^{-}_b$}
\def\Xibzstar         {$\Xi^{*0}_b$}
\def\Xibmstar         {$\Xi^{*-}_b$}
\def\Xibprime         {${\Xi'}^{(0,-)}_b$}
\def\Xibstar          {$\Xi^{*(0,-)}_b$}

\newcommand{\Photo}{}

\begin{document}
\vspace*{4cm}
\title{Heavy flavour spectroscopy at LHCb}
\author{ P. Gandini (on behalf of the LHCb collaboration) }
\address{Istituto Nazionale di Fisica Nucleare, Sezione di Milano\\
via Celoria 16, 20133, Milan, Italy.}
\maketitle\abstracts{In this talk, we present the latest experimental results on heavy flavour spectroscopy at the LHCb detector. 
The first observation of two baryonic resonances is reported in the $\Xi_b^{(-,0)} \pi^+\pi^-$ final states 
and a study of charmonium decays to $K_s^0 K \pi$ in $B \rightarrow (K_s^0 K \pi) K$ decays is presented.}

\section{Introduction}
\begin{figure}[bth]
    \centering
    \includegraphics[width=1.0\linewidth]{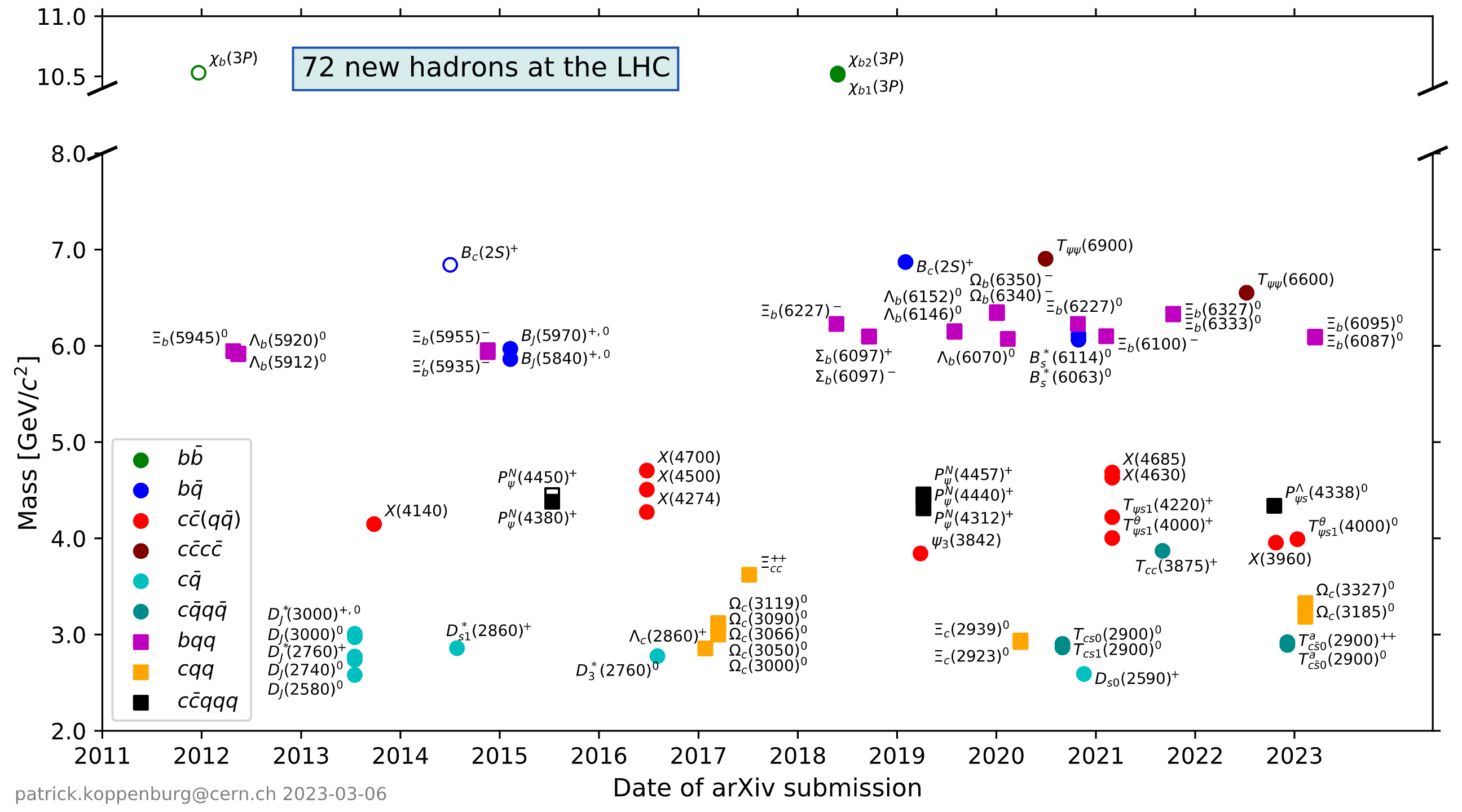}
    \caption{List of the observed hadrons at LHCb, at the time of writing.\label{fig:masses}}
\end{figure}

The LHCb experiment \cite{lhcb} is an extraordinary spectroscopy gym for both \emph{conventional} and \emph{exotic} states, given the high production cross-sections of hadronic states at the LHC.
Experimental searches of new states benefit from the excellent tracking performance, particle identification of charged particles and trigger efficiency of the LHCb detector. 
Figure~\ref{fig:masses} shows the observed hadrons at LHC since its start, at the time of writing \cite{listhadron}.

\section{Observation of new baryons in the $\Xi_b^- \pi^-\pi^+$ and $\Xi_b^0 \pi^-\pi^+$ systems}
\begin{figure}[t]
    \centering
    \includegraphics[width=0.8\linewidth]{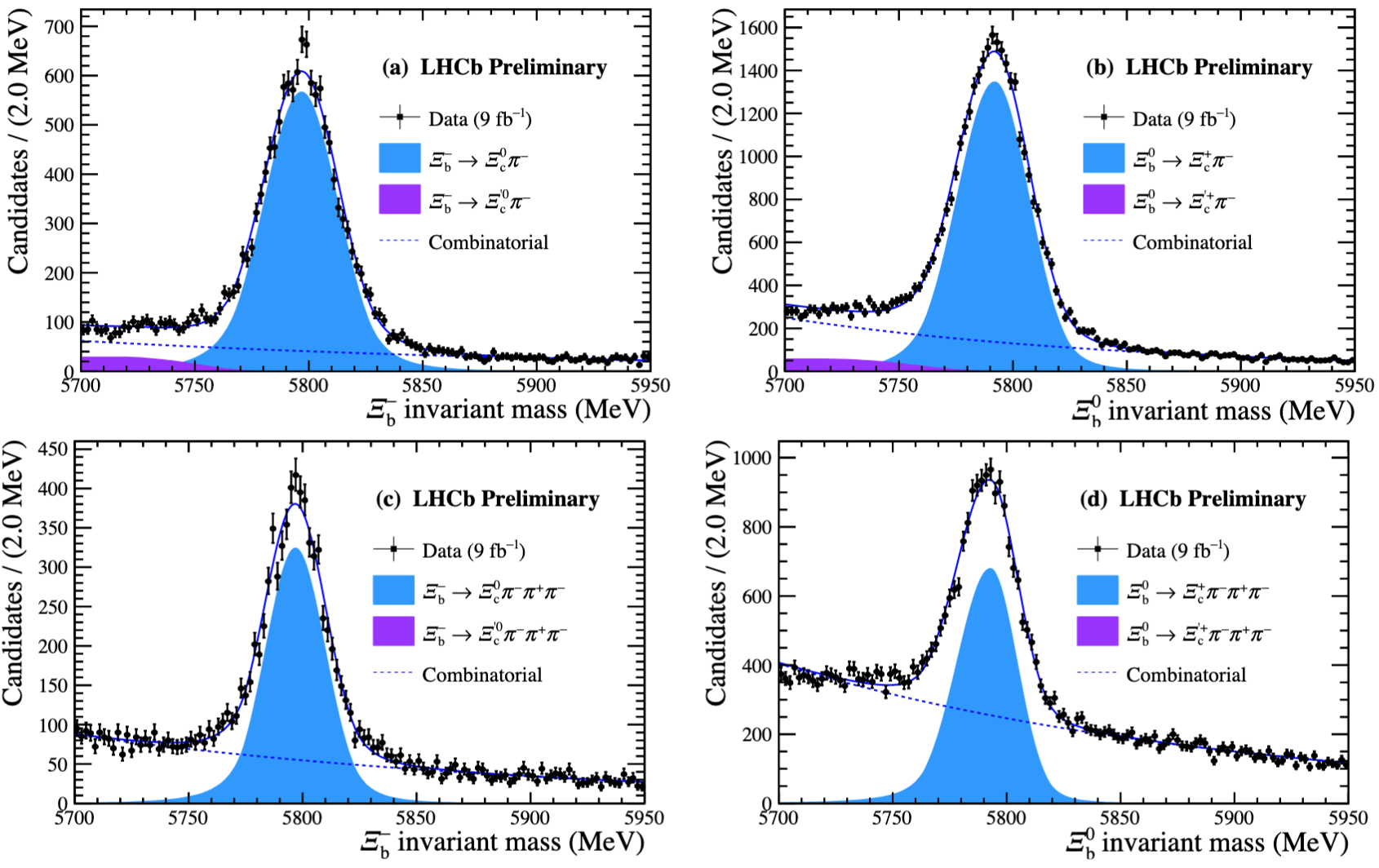}
    \caption{   
    Invariant-mass distributions of the selected signal candidates.
    $\Xi_b^- \rightarrow \Xi_c^0 \pi^-$ (a),
    $\Xi_b^0 \rightarrow \Xi_c^+ \pi^-$ (b),
    $\Xi_b^- \rightarrow \Xi_c^0 \pi^- \pi^+ \pi^-$ (c),
    $\Xi_b^0 \rightarrow \Xi_c^+ \pi^- \pi^+ \pi^-$ (d) decays.    
    \label{fig:xib}}
\end{figure}

\begin{figure}[t]
    \centering
    \includegraphics[width=1.0\linewidth]{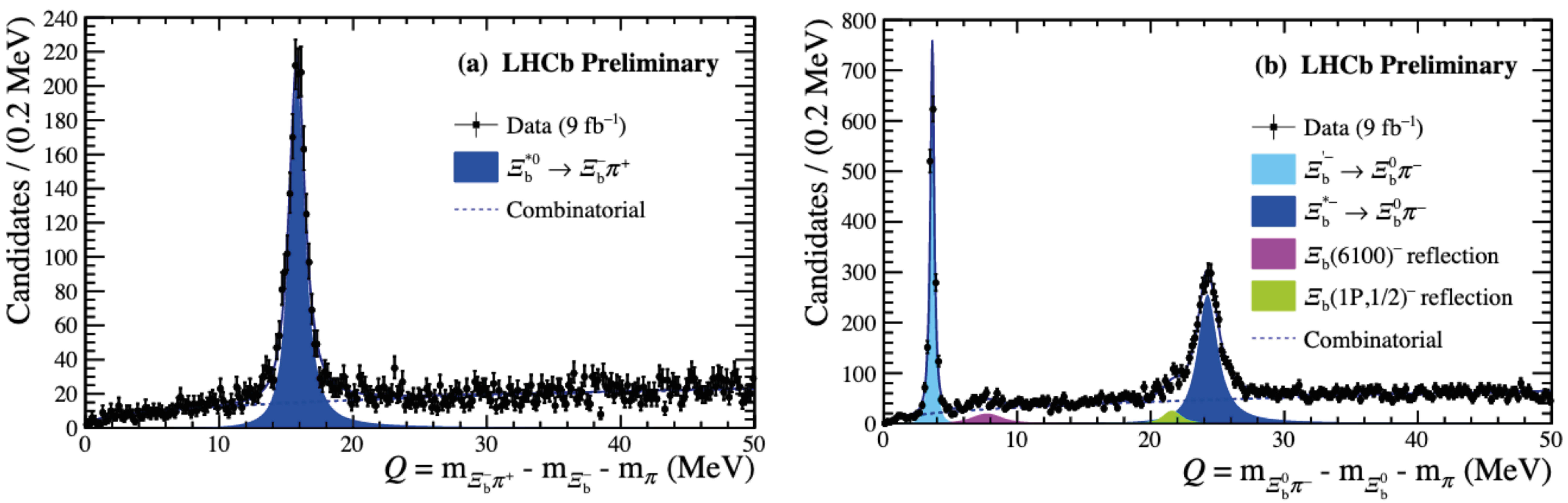}
    \caption{   
    $Q$ value distributions of selected $\Xi_b^- \pi^+$  (a) and $\Xi_b^0 \pi^-$ (b) candidates.
    \label{fig:Xibpi}}
\end{figure}

\begin{figure}[tbh]
    \centering
    \includegraphics[width=1.0\linewidth]{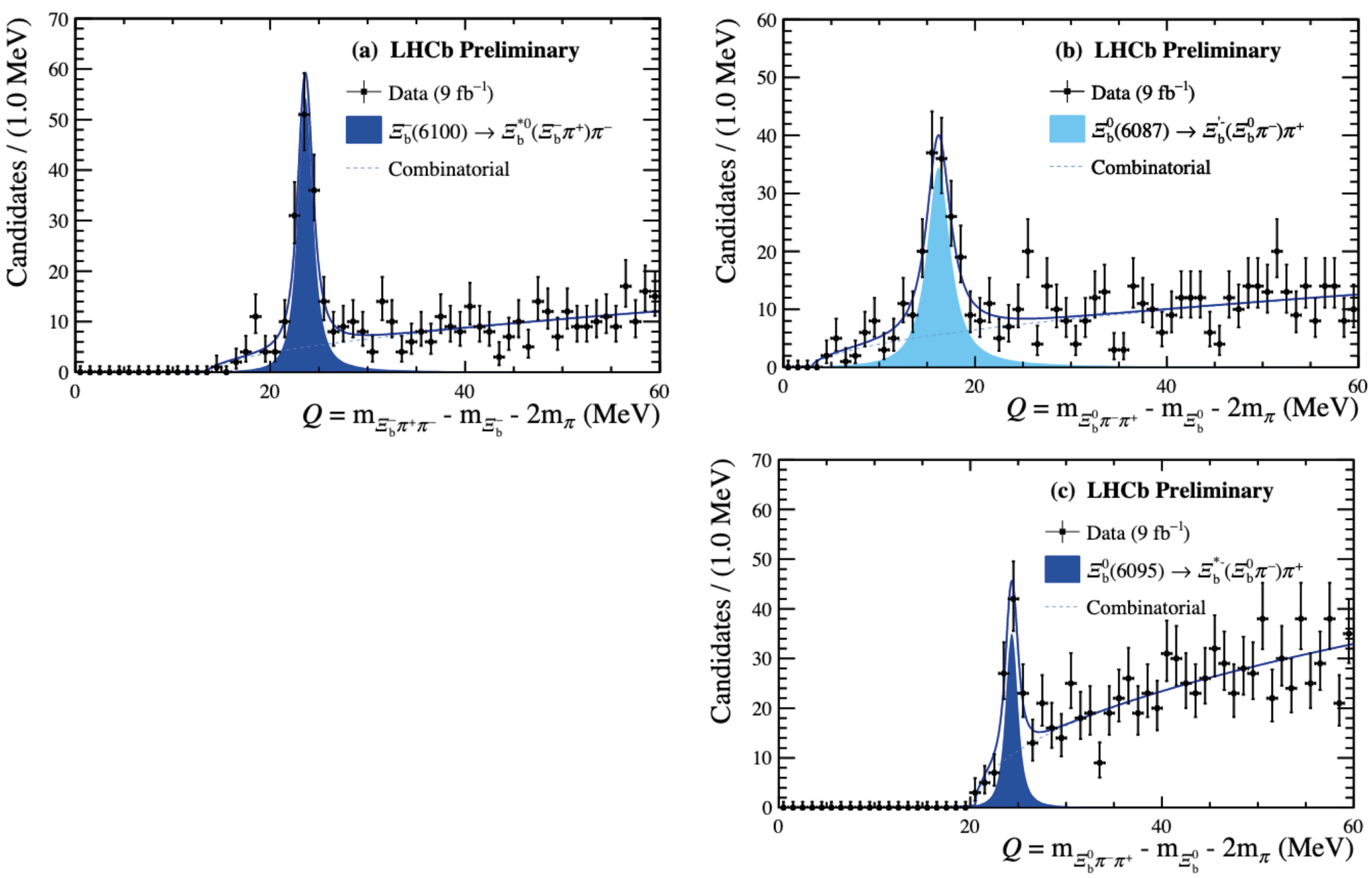}
    \caption{
    $Q$ value distributions of selected
    $\Xi_b^- \pi^+ \pi^-$ candidates in the \Xibzstar mass window (a), 
    $\Xi_b^0 \pi^- \pi^+$ candidates in the \Xibmprime mass window (b),
    $\Xi_b^0 \pi^- \pi^+$ candidates in the \Xibmstar mass window (c).
    \label{fig:Xibpipi}}
\end{figure}

We present the first investigation~\cite{LHCbPAPER2023008} at the LHCb detector of the final states $\Xi_b \pi\pi $.
The \Xibsigned baryons form an isospin doublet and are made of a $b$ quark,
an $s$ quark and a lighter $q$ ($u$ or $d$) quark.
Their ground states have angular momentum $L=0$ between the $b$ quark and the light diquark.
Three isospin doublets of such non-excited states are expected
with different spin-parity $J^P$ and light diquark angular momentum $J_{sq}$.
The \Xibsigned, \Xibprime and \Xibstar states are characterized by  ($J_{sq}$, $J^P$) values of ($0,{\frac{1}{2}}^+$), ($1,{\frac{1}{2}}^+$) and ($1, {\frac{3}{2}}^+$), respectively.
All states have been observed experimentally but one, the \Xibzprime.
The CMS collaboration has reported the observation of the \XibmstarstarB resonance in the $\Xi_b^- \pi^+ \pi^-$ final state,
using $\Xi_b^-$ decays to final states containing $J/\Psi$ mesons~\cite{PhysRevLett.126.252003}.
In this analysis, both the $\Xi_b^- \pi^+ \pi^-$ and $\Xi_b^0 \pi^+ \pi^-$ final states  
are investigated experimentally
(the inclusion of charge conjugate processes and the use of natural units are implicit),
using $pp$ collision data collected by the LHCb experiment
at $\sqrt{s} = 7,8,13$ TeV, corresponding to an integrated luminosity of 9 $\mathrm{fb}^{-1}$.
Contributions from intermediate $\Xi_b^- \pi^+$ and $\Xi_b^0 \pi^-$ resonances are also investigated.
Samples of $\Xi_b^-$ ($\Xi_b^0$) candidates are formed
from $\Xi_c^0 \pi^-$ ($\Xi_c^+ \pi^-$) and
$\Xi_c^0 \pi^- \pi^+ \pi^-$ ($\Xi_c^+ \pi^- \pi^+ \pi^-$) combinations,
where the $\Xi_c^0$ ($\Xi_c^+$) baryon is reconstructed in the $p K^- K^- \pi^+$ ($p K^- \pi^+$) final state.
\Xibzstar, \Xibmprime and \Xibmstar states are reconstructed in
$\Xi_b^- \pi^+$ and $\Xi_b^0 \pi^-$ final states.
The mass distributions of the selected $\Xi_b$ 
candidates are shown in~Figure~\ref{fig:xib}.
The $\Xi_b \pi$ mass spectra are shown in Figure~\ref{fig:Xibpi}.
The fitted yields are $2019 \pm 58$ for the \Xibzstar\ baryon,
$1750 \pm 50$ for the \Xibmprime\ baryon and $3380 \pm 110$ for the \Xibmstar\ baryon.
The \XibmstarstarB\ state is confirmed in the \Xibzstar\ $\pi^-$ mass 
distribution (Figure~\ref{fig:Xibpipi}a), while two new peaks are observed in the 
\Xibmprime $\pi^+$  (Figure~\ref{fig:Xibpipi}b)
and \Xibmstar $\pi^+$  (Fig.~\ref{fig:Xibpipi}c)
mass distributions, referred to as \XibzstarstarA\ and \XibzstarstarB.
For the newly observed states, the numbers of signal events are $136 \pm 17$ for
the \XibmstarstarB\ resonance, $147 \pm 19$ for the \XibzstarstarA\ resonance and $69 \pm 14$ for the \XibzstarstarB\ resonance.
The local significances of the three peaks are
$18 \sigma$, $15 \sigma$ and $9 \sigma$, respectively, based on the differences in
log-likelihood between a fit with zero signal and the nominal fit.
These significances are reduced to  $12 \sigma$, $10 \sigma$ and $8 \sigma$, respectively, once systematic uncertainties on the yields are taken into account in the likelihood profiles.
In summary, the observation of two new $bsq$ baryons is reported.
The observation of the charged state by the CMS collaboration, although with improved significance and sensitivity on its physical parameters. 
The two excited $\Xi_b^0$ states are observed for the first time.
This measurement uses final states with up to 9 tracks, most of which are pions, showing excellent performance of the LHCb tracking, reconstruction and particle identification systems.
Finally, the decay mode $\Xi_b^0 \rightarrow \Xi_c^+ \pi^- \pi^+ \pi^-$ is observed for the first time
and used for this search. The properties of the $\Xi_b^{*0}$, $\Xi_b^{'-}$ and $\Xi_b^{*-}$
baryons are measured with high precision.
Data indicate that the \XibmstarstarB\ baryon decays mainly through the \Xibzstar\ $\pi^-$ state, while the \XibzstarstarA\ baryon mainly through \Xibmprime $\pi^+$ state and the \XibzstarstarB\ baryon mainly through \Xibmstar $\pi^+$ state,
with no significant contributions to the signals from events
outside their respective $m_{\Xi_b \pi}$ mass windows.
A naive interpretation would be that the new states are $P$-wave states
($l=1$ between $b$ and $qs$ diquark)
coupling to the $b$ quark
to give a pair of states with $J^P = {\frac{1}{2}}^-$ and
${\frac{3}{2}}^-$.

\section{Study of charmonium decays to $K_s^0 K \pi$ in $B \rightarrow (K_s^0 K \pi) K$}
\begin{figure}[tbh]
    \centering
    \includegraphics[width=1.0\linewidth]{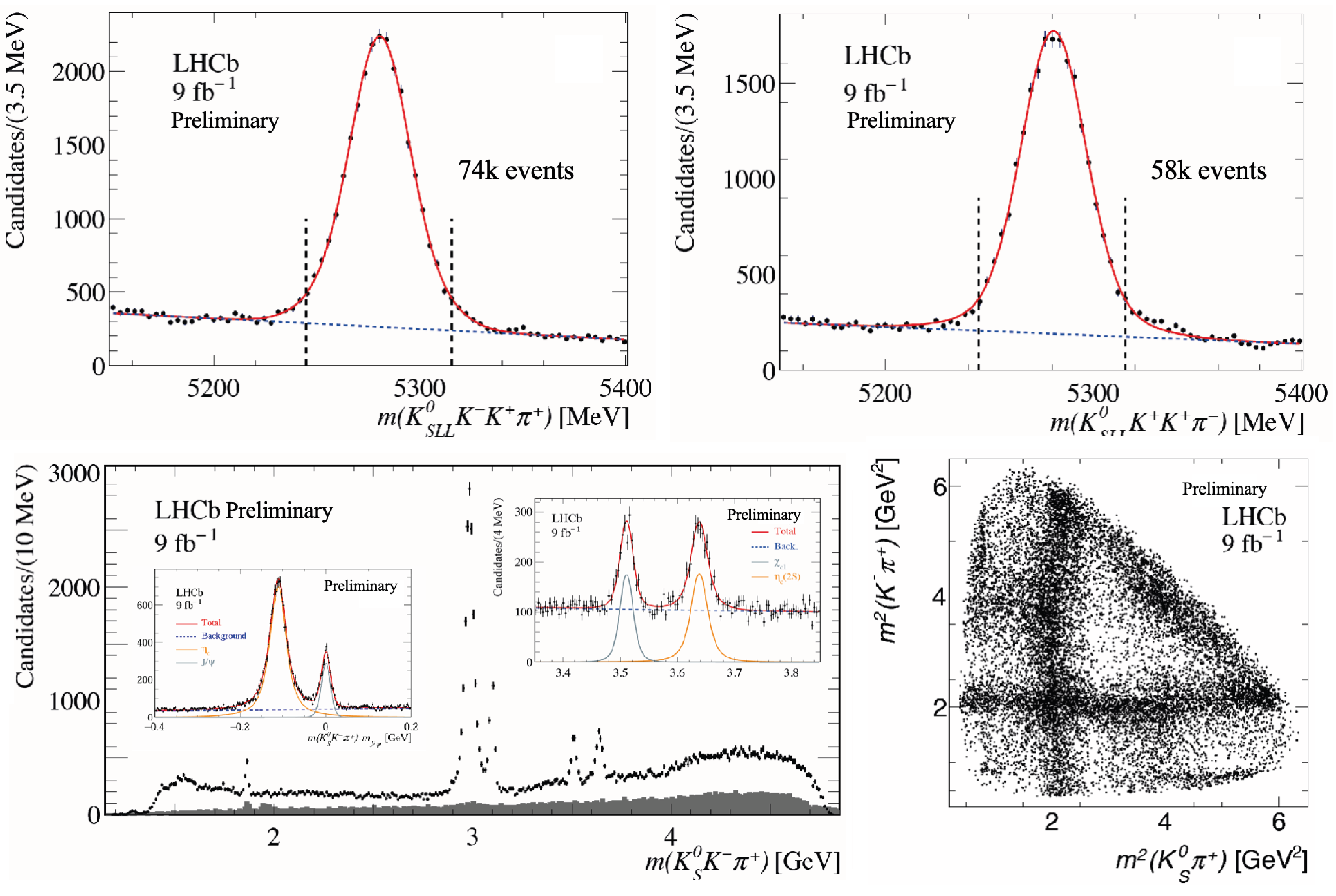}
    \caption{Top: signal mass plots and yields for the $B \rightarrow (K_s^0 K \pi) K$ decays. Bottom: different views of the $K_s^0 K^- \pi^+$ distribution with zoomed details on the interesting regions.
    The Dalitz plot of $\eta_c \rightarrow K_s^0 K \pi$ is also shown.\label{fig:dalitz}}
\end{figure}

In this very detailed paper~\cite{LHCbPAPER2022051}, we report a high-statistics study of 
the $B^+ \rightarrow K_s^0 K^+ K^- \pi^+$ and $B^+ \rightarrow K_s^0 K^+ K^+ \pi^-$.
The invariant mass spectra of both $B^+$ decay modes reveal a rich content of charmonium resonances in 
$K_s^0 K \pi$. A sketch of the invariant mass spectra considered in the analysis is shown in Figure \ref{fig:dalitz}. New precise measurements of the $\eta_c$ and $\eta_c(2S)$ resonance parameters
are performed and branching fraction measurements are obtained for $B^+$ decays to 
$eta_c$, $J/\Psi$, $\eta_c(2S)$ and $\chi_{c1}$ resonances.
In particular, the first observation and branching fraction measurement of
$B^+ \rightarrow \chi_{c0} K_s^0 \pi^+$ is reported as well as first measurements of
$B^+ \rightarrow K_s^0 K^+ K^- \pi^+$ and $B^+ \rightarrow K_s^0 K^+ K^+ \pi^-$ branching fractions.
Dalitz plot analyses of $\eta_c \rightarrow K_s^0 K \pi$ and $\eta_c(2S) \rightarrow K_s^0 K \pi$
decays are performed.
A new measurement of the amplitude and phase of the $K \pi$ S-wave as functions of the $K \pi$
mass is performed,
together with measurements of the $K^{*0}(1430)$, $K^{*0}(1950)$ and $a_0(1700)$ parameters.
Finally, the branching fractions of $\chi_{c1}$ decays to $K^*$ resonances are also measured.


\end{document}